\documentclass[11pt]{article}
\usepackage{authblk}
\usepackage[backend = biber,style = apa,block=ragged]{biblatex}
\usepackage{graphicx}
\usepackage{xcolor}
\usepackage{hyperref}
\usepackage{amsmath}
\usepackage{tikz}
\usepackage[margin=2.54cm]{geometry}
\usetikzlibrary{arrows.meta,calc,positioning,decorations.pathreplacing}

\usepackage{caption}

\definecolor{linkred}{HTML}{C0392B}
\definecolor{citegreen}{HTML}{1E8449}
\definecolor{citeslate}{HTML}{34527A}
\definecolor{urlteal}{HTML}{0891A6}

\hypersetup{
    colorlinks=true,
    linkcolor=linkred,
    citecolor=citegreen,
    filecolor=citeslate,
    urlcolor=urlteal,
    }

\title{A place for stabilization alongside tipping cascades:\\the AMOC--cryosphere system}
\author[1,2]{Sacha Sinet\thanks{Email: s.a.m.sinet@uu.nl}}
\affil[1]{Department of Physics, Institute for Marine and Atmospheric Research Utrecht, Utrecht, the Netherlands}
\affil[2]{Center for Complex Systems Studies, Utrecht University, Utrecht, the Netherlands}

\date{}

\begin{document}
\maketitle

\begin{abstract}
The Atlantic Meridional Overturning Circulation (AMOC) and cryosphere are central to tipping-point research, where their interactions are typically framed as pathways for cascading transitions. Here, I review evidence that these interactions can also oppose change, affecting its magnitude and timing or, in some cases, whether a transition occurs at all. I argue that a more complete assessment of tipping risk requires progress on three fronts: systematic attention to stabilizing alongside destabilizing dynamics, conceptual frameworks that accommodate competing and context-dependent effects, and methods to detect, anticipate, and attribute stabilization.
\end{abstract}

\section{From tipping to cascades: a framework of instability}

Anthropogenic climate change is already reshaping the Earth system, increasingly threatening the stability of the climatic conditions on which present ecosystems and human societies depend \parencite{IPCC_2021_WGI_SPM}. While this alone provides ample reason for rapid and sustained climate action, some Earth-system components may additionally undergo abrupt, self-sustaining, and potentially irreversible change once critical thresholds, or tipping points, are crossed \parencite{lentonTippingElementsEarths2008}. These tipping elements, among which the Atlantic Meridional Overturning Circulation (AMOC) is one of the most prominent, add a further layer of risk and uncertainty to an already dangerous climate trajectory.

Crucially, tipping elements are not independent. Changes in one can alter the stability of others, creating pathways through which tipping risk reverberates across the Earth system \parencite{wunderlingClimateTippingPoint2024}. The AMOC and cryosphere provide a particularly relevant example, forming a tightly coupled system that includes several core tipping elements, notably the AMOC, Greenland Ice Sheet (GIS), and the West Antarctic Ice Sheet (WAIS) \parencite{armstrongmckayExceeding15degCGlobal2022}. In the canonical pathway, warming increases GIS melt, freshening the North Atlantic and potentially weakening the AMOC, an effect that can be amplified by the salt-advection feedback \parencite{weijerStabilityAtlanticMeridional2019}. The resulting reduction in northward heat transport redistributes heat toward the Southern Hemisphere, promoting basal melting of the WAIS, where marine ice-sheet instability can further amplify retreat \parencite{schoofIceSheetGrounding2007}. In the worst case, tipping of one element can trigger others, like falling dominoes, producing a tipping cascade \parencite{dekkerCascadingTransitionsClimate2018,kloseWhatWeMean2021}.

This tipping-centered framing has been scientifically productive, but it naturally gives greater prominence to processes that push systems toward instability than to those that oppose, delay, or limit change. This is especially apparent in work on tipping cascades, where the central question becomes how interactions transmit or amplify tipping risk. In this commentary, I focus instead on stabilizing dynamics, using the AMOC--cryosphere system as a case study. I review evidence for such dynamics operating within and between its components, and argue that bringing them more systematically into the picture is necessary for a more complete assessment of tipping risks.

\section{The case for stabilizing dynamics}

\begin{figure}[htbp]
\centering
\begin{tikzpicture}[
  domino/.style={fill=gray!10, draw=black!50, thick, rounded corners=1pt},
  dominostable/.style={fill=urlteal!45, draw=urlteal!85!black, thick, rounded corners=1pt},
  ghost/.style={fill=gray!6, draw=gray!45, thick, dashed, rounded corners=1pt},
  motion/.style={->, thick, gray!65, shorten >=2pt, shorten <=2pt},
  base/.style={gray!30, thin},
  tag/.style={font=\small\bfseries, gray!45!black},
  every node/.style={font=\small}
]

\begin{scope}[shift={(0,0)}]
  \draw[base] (-0.35,0) -- (2.55,0);
  \begin{scope}[shift={(0,0)}, rotate=-68]
    \draw[domino] (-0.3,0) rectangle (0,1.1);
  \end{scope}
  \begin{scope}[shift={(1.1,0)}, rotate=-35]
    \draw[domino] (-0.3,0) rectangle (0,1.1);
  \end{scope}
  \begin{scope}[shift={(2.2,0)}, rotate=0]
    \draw[domino] (-0.15,0) rectangle (0.15,1.1);
  \end{scope}
  \draw[motion] (-0.55,0.85) to (-0.2,0.85);
  \node[tag] at (1.1,-0.6) {Cascading tipping};
\end{scope}

\begin{scope}[shift={(4.6,0)}]
  \draw[base] (-0.35,0) -- (2.55,0);
  \begin{scope}[shift={(0,0)}, rotate=-35]
    \draw[domino] (-0.3,0) rectangle (0,1.1);
  \end{scope}
  \begin{scope}[shift={(1.1,0)}, rotate=0]
    \draw[dominostable] (-0.15,0) rectangle (0.15,1.1);
  \end{scope}
  \begin{scope}[shift={(2.2,0)}, rotate=35]
    \draw[domino] (0,0) rectangle (0.3,1.1);
  \end{scope}
  \draw[motion] (-0.55,0.85) to (-0.2,0.85);
  \node[tag] at (1.1,-0.6) {Opposing forces};
\end{scope}

\begin{scope}[shift={(9.6,0)}]
  \draw[base] (-0.35,0) -- (1.5,0);
  \begin{scope}[shift={(0,0)}, rotate=-40]
    \draw[domino] (-0.3,0) rectangle (0,1.1);
  \end{scope}
  \begin{scope}[shift={(1.1,0)}, rotate=-40]
    \draw[ghost] (-0.3,0) rectangle (0,1.1);
  \end{scope}
  \begin{scope}[shift={(1.1,0)}, rotate=0]
    \draw[dominostable] (-0.15,0) rectangle (0.15,1.1);
  \end{scope}
  \draw[motion] (-0.55,0.85) to (-0.2,0.85);
  \node[tag] at (0.5,-0.6) {Restoring force};
\end{scope}

\begin{scope}[shift={(13.9,0)}]
  \draw[base] (-0.35,0) -- (1.0,0);
  \begin{scope}[shift={(0,0)}, rotate=0]
    \draw[dominostable] (-0.15,0) rectangle (0.15,1.1);
  \end{scope}
  \begin{scope}[shift={(0.65,0)}, rotate=0]
    \draw[dominostable] (-0.15,0) rectangle (0.15,1.1);
  \end{scope}
  \draw[urlteal!85!black, very thick] (0.15,0.55) -- (0.5,0.55);
  \draw[motion] (-0.55,0.85) to (-0.2,0.85);
  \node[tag] at (0.325,-0.6) {Mutual stabilization};
\end{scope}

\end{tikzpicture}
\captionsetup{width=0.9\textwidth}
\caption{\emph{Dominoes naturally represent the propagation of instability following an initial perturbation (gray arrow), but do not naturally accommodate other forms of interaction, such as opposing influences, restoration, or mutual stabilization. Cascading tipping can also take many forms beyond the one shown here \parencite{kloseWhatWeMean2021}.}}
\label{fig:domino}
\end{figure}
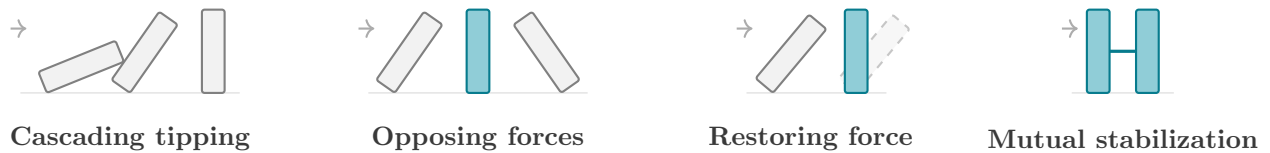

The dominant metaphor for tipping-element interactions is the domino: push one, and it knocks over the next, reinforcing the ``Hothouse Earth'' narrative \parencite{steffenTrajectoriesEarthSystem2018}. But dominoes naturally represent a linear, one-way chain of cause and effect rather than multicausal interactions and recursive feedbacks \parencite{lawrenceGlobalPolycrisisCausal2024}. They are therefore less suited to depicting interactions that buffer change (Figure~\ref{fig:domino}), shaping intuition more naturally around cascading instability than stabilizing dynamics.

However, the Earth system is not characterized by instability alone, as illustrated by the comparatively stable conditions of the Holocene, which began approximately 11,700 years ago near the end of the last deglacial transition \parencite{IPCC_2021_WGI_Ch_2}. At a more fundamental level, the strongest individual radiative feedback in the present climate, the Planck response, is stabilizing: warming increases outgoing longwave radiation and acts to restore radiative balance \parencite{IPCC_2021_WGI_Ch_7}. Stabilizing dynamics can likewise operate within tipping elements and across their interactions with other parts of the Earth system. Here, I use the term stabilizing more broadly than in its strict dynamical sense to refer to processes, feedbacks, or interactions that oppose a specified change. These can take distinct forms: buffering a response by reducing its magnitude or rate, providing a restoring feedback that counteracts a perturbation, and delaying or preventing a transition.

The Northern Hemisphere cooling associated with AMOC weakening provides several examples of stabilizing effects. Simulations by \textcite{leeWeakenedAtlanticMeridional2023} show that AMOC weakening reduced the 1980--2020 decline rates of Arctic sea-ice area and volume by 36\% and 22\%, respectively. AMOC-induced cooling can also substantially reduce permafrost carbon loss during a temporary AMOC slowdown, by 26--57\% across the experiments of \textcite{steinertPermafrostCarbonRelease2026}. For the GIS, coupled simulations show that cooling following an AMOC collapse can effectively halt GIS melting at ${\sim}3^\circ$C global warming and substantially delay disintegration at higher warming levels \parencite{poppelmeierMutualStabilizationAMOC2025}.

The same reduction in northward heat transport also drives the thermal bipolar seesaw \parencite{crowleyNorthAtlanticDeep1992}, producing Southern Hemisphere warming that would generally be expected to increase mass loss from the Antarctic Ice Sheet, particularly the WAIS. Yet the response around the Antarctic Ice Sheet is more complex, with subsurface waters near the continental margin able to cool despite surface warming \parencite{berdahlAntarcticClimateResponse2024}. A recent model study accordingly finds little response in total Antarctic ice volume for several centuries after AMOC shutdown and no destabilization of the WAIS, with longer-term subsurface cooling instead reducing basal melt \parencite{hoseSimulatingImpactAMOC2026}.

The cryosphere can, in turn, influence AMOC stability through meltwater forcing. Greenland meltwater provides the familiar destabilizing pathway \parencite{bakkerFateAtlanticMeridional2016}, whereas the influence of Antarctic meltwater is less straightforward, partly because competing processes act on different timescales \parencite{swingedouwImpactTransientFreshwater2009}. Model studies have reported AMOC responses to Antarctic freshwater input ranging from slight strengthening \parencite{liGlobalClimateImpacts2023} or weakening \parencite{seidovThereSimpleBipolar2005,stoufferClimateResponseExternal2007}, to delayed weakening \parencite{sadaiFutureClimateResponse2020} and even re-establishment of overturning from an AMOC off state \parencite{weaverMeltwaterPulse1A2003}. More recent work shows that these contrasting responses can also coexist within a single model framework: varying the relative timing and duration of ice-sheet freshwater input can increase or decrease AMOC resilience in several different ways, including cases in which meltwater from WAIS tipping prevents an AMOC collapse \parencite{sinetMeltwaterWestAntarctic2025}.

Stabilizing processes also operate within individual subsystems. As the AMOC weakens, reduced northward heat transport provides a well-established thermal-advection feedback opposing further weakening \parencite{rahmstorfRoleTemperatureFeedback1995, garubaAtlanticMeridionalOverturning2025}. A recent CESM feedback decomposition also shows a stabilizing contribution from gyre freshwater transport and a salt-advection feedback that switches between stabilizing and destabilizing depending on the background freshwater transport \parencite{vanderborghtFeedbackProcessesCausing2025}. Ice sheets likewise contain feedbacks that can damp their response to forcing \parencite{fykeOverviewInteractionsFeedbacks2018}. Among them, glacial isostatic adjustment and the associated bedrock uplift can weaken the melt--elevation feedback in Greenland \parencite{zeitzDynamicRegimesGreenland2022} and delay grounding-line retreat in Antarctica by 50--130 years in the simulations of \textcite{vancalcarBedrockUpliftReduces2025}, while reducing the Antarctic sea-level contribution by 9--23\% by 2500. Ice loss also lowers local sea level through reduced gravitational attraction, further reinforcing stabilization \parencite{gomezSeaLevelStabilizing2010}.

Taken together, these examples show that stabilizing dynamics are a recurring feature of the AMOC--cryosphere system, operating through multiple mechanisms both within individual components and across their interactions. Their effects can depend on system state, forcing, and timescale, and can shape the extent and pace of change and, in some cases, whether a transition happens at all. Although not considered here, similar dynamics also arise in other tipping-element interactions, a good example being the potentially stabilizing effects of AMOC collapse on parts of the Amazon rainforest \parencite{nianPotentialCollapseAtlantic2023}. I make no claim here about whether stabilizing effects generally outweigh or are outweighed by destabilizing influences. The more basic point is that stabilizing dynamics are sufficiently recurrent and consequential to require systematic integration into the study and assessment of tipping risks.

\section{Towards a more complete assessment of tipping risks}

\begin{figure*}[ht]
\centering
\includegraphics[width=0.95\textwidth]{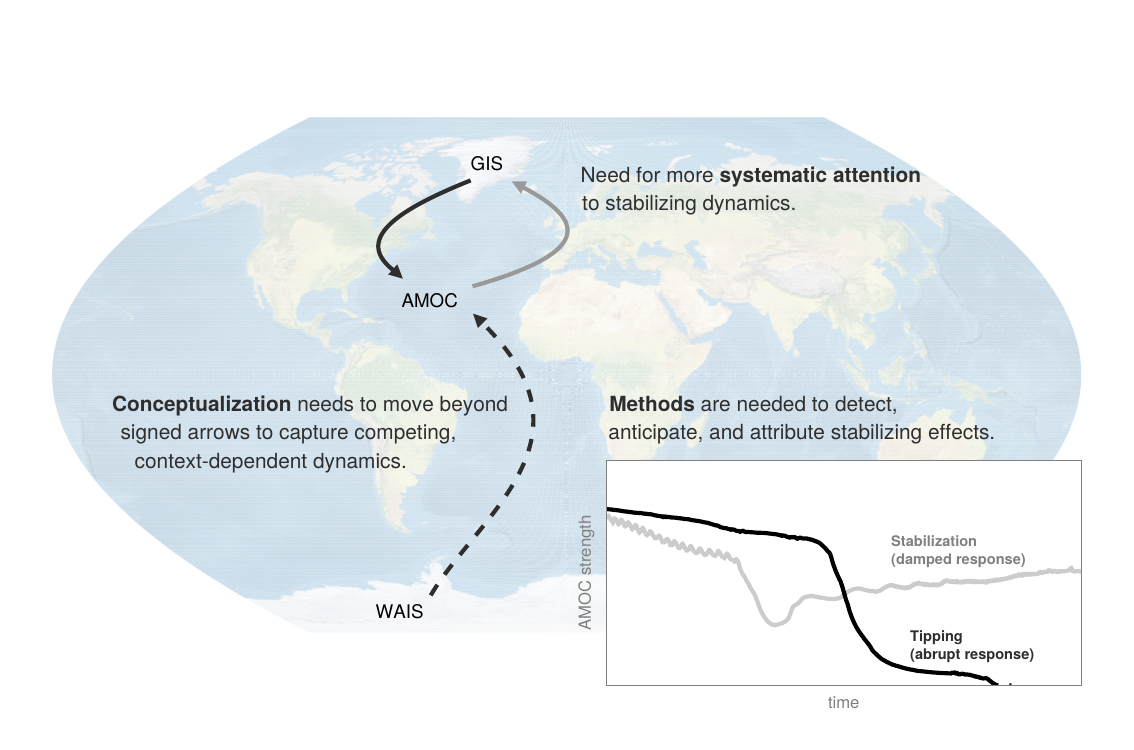}
\captionsetup{width=0.9\textwidth}
\caption{
\emph{Three areas requiring particular attention, illustrated through AMOC--ice sheet interactions. Arrow prominence highlights the need for more systematic attention to stabilizing dynamics, while the dashed arrow highlights the need for conceptualization to move beyond signed arrows and capture competing, context-dependent dynamics. The inset shows contrasting stabilizing and tipping responses from the experiments of \textcite{sinetMeltwaterWestAntarctic2025}.}}
\label{fig:gaps}
\end{figure*}

A complete assessment of tipping risks requires systematic treatment of both stabilizing and destabilizing dynamics. In the most recent assessment of tipping-element interactions, 15 out of 20 were classified as destabilizing, compared with four as stabilizing and one with an unclear or competing sign \parencite{armstrongmckayStatusEarthSystem2025}. This imbalance may well reflect current knowledge, and the classifications are mostly consistent with the AMOC--cryosphere evidence reviewed above. What is less clear is how such an imbalance should be interpreted, including whether stabilizing and destabilizing influences have been examined on comparable terms. I highlight three areas where this question requires particular attention: research focus, conceptual framing, and methodology, as illustrated in Figure~\ref{fig:gaps}.

\paragraph{Systematic attention to stabilizing dynamics:} Tipping research has understandably focused on pathways that propagate instability, naturally placing less emphasis on stabilizing feedbacks and interactions. One particularly illustrative example is the AMOC--GIS coupling, where the stabilizing AMOC$\rightarrow$GIS pathway appears to have received less direct attention than the familiar destabilizing GIS$\rightarrow$AMOC pathway, even though the two are assessed as comparable in strength \parencite{wunderlingClimateTippingPoint2024}. An important exception is work on safe overshoot \parencite{ritchieImplicationsOvershooting152026}, where tipping avoidance, including the avoidance of tipping cascades \parencite{sinetCriterionSafeOvershoot2026}, is itself the object of study. The point, then, is not whether stabilizing dynamics exist, but whether they are being considered as systematically as destabilizing ones.

\paragraph{Conceptualizing competing interactions:} Representing tipping-element interactions as signed arrows between climate subsystems is useful for synthesis, but necessarily compresses the mechanisms and conditions that determine their effect. Recent tipping-element assessments make this simplification explicitly \parencite{wunderlingClimateTippingPoint2024,armstrongmckayStatusEarthSystem2025}, while also acknowledging that interaction signs can depend on multiple processes, system state, and spatio-temporal scales. The WAIS$\rightarrow$AMOC interaction is a clear example: West Antarctic meltwater forcing can promote AMOC tipping or recovery under some conditions, with opposite effects emerging under others. A single signed link can therefore mask competing influences whose net effect depends on context, suggesting that interactions should also be characterized by the mechanisms and conditions that determine their sign.

\paragraph{Detecting, anticipating, and attributing stabilization:} Stabilizing effects can be particularly elusive because, when successful, they may manifest through what does not happen, a familiar challenge across domains concerned with prevention and risk reduction \parencite{finebergParadoxDiseasePrevention2013,joreOntologicalEpistemologicalChallenges2021,hasselFrameworkEvaluatingSocietal2021}. At the same time, methodological development has largely centered on tipping itself, with methods specifically tailored to detecting and anticipating transitions, including early-warning indicators of declining resilience \parencite{schefferEarlywarningSignalsCritical2009,dakosTippingPointDetection2024} and methods for identifying abrupt shifts \parencite{terpstraAssessmentAbruptShifts2025a}. Assessing stabilizing dynamics on comparable terms will require similarly dedicated tools to detect and anticipate their effects and attribute them to underlying processes. Promising directions include causal-inference methods, which can help isolate causal pathways in complex Earth-system data \parencite{rungeInferringCausationTime2019}, and physics-based feedback decompositions, which explicitly separate stabilizing and destabilizing contributions where the underlying physics is sufficiently understood \parencite{vanderborghtFeedbackProcessesCausing2025}.

Ultimately, broadening the picture is not about downplaying cascading tipping risks, but about assessing future climate risks more completely. Tipping points sharpen the case for mitigation because delaying action increases the risk of crossing thresholds beyond which change may become increasingly difficult to manage \parencite{mollerAchievingNetZero2024}. Stabilizing dynamics may nevertheless arise as unintended consequences of a changing climate and alter which risks unfold, when, and to what extent. Addressing these three areas would help bring such effects more systematically into the picture, not to reduce urgency, but to better characterize the climate risks we may ultimately need to adapt to.

\subsection*{Declaration of conflicting interests} The author declares no conflicts of interest.
\subsection*{Data availability} No new data were generated for this commentary. The model output shown in Figure~\ref{fig:gaps} is taken from \textcite{sinetMeltwaterWestAntarctic2025}, where the corresponding data are publicly available.
\subsection*{Funding} This project has received funding from the Dutch Research Council (NWO) through the NWO-Vici
project ``Interacting tipping elements: When does tipping cause tipping'' (Project No. VI.C.202.081).

\printbibliography

\end{document}